# Stochastic Memristive Devices for Computing and Neuromorphic Applications

Siddharth Gaba, Patrick Sheridan, Jiantao Zhou, Shinhyun Choi and Wei Lu*

Department of Electrical Engineering and Computer Science, University of Michigan, MI, 48109, USA

*Corresponding author. Electronic mail: wluee@eecs.umich.edu

*Abstract*— **Nanoscale resistive switching devices (memristive devices or memristors) have been studied for a number of applications ranging from non-volatile memory, logic to neuromorphic systems. However a major challenge is to address the potentially large variations in space and in time in these nanoscale devices. Here we show that in metal-filament based memristive devices the switching can be fully stochastic. While individual switching events are random, the distribution and probability of switching can be well predicted and controlled. Rather than trying to force high switching probabilities using excessive voltage or time, the inherent stochastic nature of resistive switching allows these binary devices to be used as building blocks for novel error-tolerant computing schemes such as stochastic computing and provide a needed "analog" feature in neuromorphic applications. To verify such potential, we demonstrated memristor-based stochastic bitstreams in both time and space domains, and show that an array of binary memristors can act as a multi-level "analog" device for neuromorphic applications.**

With conventional transistors approaching their scaling limit, much attention has been focused on emerging device concepts (1). Devices based on resistive switching (memristive devices or memristors (2-6) have been identified as a major contender in applications ranging from nonvolatile memory storage, logic, to neuromorphic systems due to their scalability (7), low power operation (8), and large connectivity offered in a crossbar structure (9). Active circuits based on memristors have also been demonstrated recently (10,11). However, a major roadblock that needs to be overcome before potential commercialization of these devices is the potentially large variations in the switching parameters (12,13). In spite of the rapid progress in the field, devices based on thin metallic electrodes and amorphous or oxide switching layers suffer from higher failure rates compared to conventional crystalline silicon devices. While some issues like *spatial variations* related to line edge roughness and film thickness irregularity can be corrected to some extent through tight process control and variation-aware design, a more challenging problem with these resistive switches is the significant randomness or *temporal variations* experienced during normal operations (9,13). This is particularly evident in devices based on metal ion migration within an insulating matrix due to the filamentary nature of the conductive path formation (14). For memory applications that require very low failure rates, excessively high programming voltages and long pulse durations are needed to ensure the switching of the devices, which causes reliability concerns. Here we show that the inherent randomness of the switching behavior (the temporal variations) can be well characterized and controlled, and that the stochastic, uncorrelated switching events can be used for novel computing schemes including stochastic computing and neuromorphic applications.

Memristive devices based on cation migration inside a solid-state electrolyte are used in this study, but the results can be generalized to other cation (5) or anion (e.g. oxygen vacancy in oxides) (6) based memristive systems as well. The device, fabricated in a two terminal crossbar form, has an amorphous silicon layer sandwiched between a silver top electrode and a polysilicon bottom electrode (9). Binary resistive switching can be reliably observed in such devices, as shown in the inset to Fig. 1a and as discussed earlier (15). It can be seen from DC sweeps that the devices exhibit standard switching hysteresis with clearly defined ON and OFF states and an apparent threshold voltage of ca. 5 V (Fig. 1a inset). During programming the voltage was applied to the top electrode while the bottom electrode was grounded. The change from OFF to ON state is explained by the formation of a dominating Ag filament (15,16). However, careful examination of the switching effects reveals that significant randomness can exist, even in the same device. The random switching effect can be prominently seen by applying bias voltages much lower than the threshold voltage of the device. For example, in Fig. 1a-c, the device was initially reset to the OFF state and a constant voltage of 2.5 V, 3.5 V and 4.5 V was applied to the device at time t = 0, respectively. The current through the device was then monitored continuously. After an initial wait time, a sharp increase in current marks the transition of the device to the ON state (e.g. Fig. 1b inset). This wait time measures the time elapsed between the application of the voltage and the switching of the device (17). Once the device was verified to have switched to the ON state, the voltage bias was turned off and the device was reset to the OFF state. This process was repeated one hundred times at each bias condition to analyze the temporal variations of the switching behavior.

The wait time before switching shows apparent randomness. For example, Fig. 1a plots the histogram for the wait times associated with an applied voltage of 2.5 V. As can be seen, even for a given voltage applied to the same device, the wait time is not fixed but rather shows a large distribution. Previous studies on such devices (4,14,17) have shown that the resistance switching is associated with the formation and rupture of a single dominant, nanoscale conducting filament. Filament formation involves oxidation, ion transport and reduction processes, all of which are thermodynamically driven (5,18) and require overcoming specific activation energies. Typically one of the processes is rate-liming so switching is associated with thermal activation over a dominant energy barrier and is thus probabilistic in nature, if only a dominant filament is involved (14). As a result, even for the same filament in the same device, the wait time will be broadly distributed and in principle can only be predicted in terms of statistics while the individual switching events occur randomly in nature.

Mathematically, if only one dominant energy barrier limits the switching process, the wait time is expected to follow a Poisson distribution and the probability that a switching event occurs within Δt at a given time t is given by

$$P(t) = \frac{\Delta t}{\tau} e^{-t/\tau} \qquad (1)$$

where τ is the characteristic wait time (14).

Fig. 1 shows that excellent fit of the wait times to the Poisson distribution in Eq. (1) can be obtained with just one fitting parameter (τ), in agreement of the hypothesis of thermodynamic activation over a dominant energy barrier during filament formation in these devices. The Poisson distribution of the wait time, with the standard deviation equaling the mean, further verifies that the switching is random and stochastic in nature.

To improve the reliability of these intrinsically non-deterministic devices for applications in deterministic storage and logic applications, feedback schemes that check the state of the device after every write operation (19), or error-control coding and redundancy (20,21) can be employed. Alternatively, excess programming voltage and long pulse width can be used to ensure the correctness of each write. By integrating Eq. (1), we see that if the programming voltage is applied for time > 5τ, then the switching probability reaches > 99%, i.e. high programming success rates can be obtained. Since the characteristic wait time τ is strongly voltage dependent, the average wait time can be reduced drastically at higher voltages too. As can be seen from Fig. 1b and Fig. 1c, the switching time distribution at different programming voltages preserves the Poisson nature but the characteristic wait time decreases significantly as the bias voltage is increased. With an increase in 2 V in the applied voltage, the characteristic time drops sharply by almost three orders of magnitude (Fig. 1d). The strong voltage dependence of (average) switching time is expected within the filament formation picture since the energy barriers for both the oxidation and the ion transport processes are field-dependent (the reduction processes of the ions is not thought to be the rate-limiting process) (5,18) and the effective barrier height is reduced upon the application of the bias voltage (14,22,23). Thus by increasing the programming voltage, τ can be reduced significantly down to the nanosecond regime (14) so switching with high success rate can be achieved. However, using excessively high voltage and excessively long pulses can result in an increased power envelope and lead to unnecessary device degradation. Instead of trying to force these non-deterministic devices to function deterministically, we show that useful functions may be obtained by taking advantage of this inherent stochastic switching nature at low voltages and shorter times.

First, we note that the switching probability can be calculated by integrating the Poisson distribution Eq. (1), which leads to

$$C(t) = 1 - e^{-t/\tau} \qquad (2)$$

For an applied voltage of 2.5 V, the prediction based on (2) is shown in Fig. 2a as the solid line. The switching probabilities can also be obtained by calculating the cumulative probability distribution function from data in Fig. 1a, shown as the squares in Fig. 2a. Again good agreements can be obtained, in agreement with the model.

The significance of Eq. (2) is that we can now predict the switching *probability* at a given programming voltage and pulse width, even though each switching event is random. We verify these predictions by applying a programming pulse with fixed amplitude (e.g. 2.5 V) and pulse width (e.g. 300 ms) and measuring whether the device was switched to the ON state during the pulse or not. The device currents measured after application of the programming pulse are shown in Fig. 2c for fifty such attempts. In this experiment, after each programming pulse the device was reset to the same original OFF state so each programming pulse starts with the same initial condition. The device switches to the ON state twenty times out of the fifty trials or attempts. This number closely matches that predicted from the theoretical curve given by Eq. (2). Obviously, longer pulses result in a higher probability of the device switching during the programming pulse, as predicted by Eq. (2). Results for 2.5 V/1000 ms pulses are shown in Fig 2d; the device switches thirty eight times compared to twenty times in Fig. 2c. Results from other pulse widths are shown in Fig 2b.

We argue this ability to predict the probability of switching in stochastic memristive switching events can be used for novel computing schemes. In one such example, the measured current can be digitized – for example, each occurrence above 0.1 µA can be assigned as "1" while each value below 0.1 µA can be assigned as "0". Thus, here we essentially obtained a stream of bits (a bitstream) which is fifty bits long in time and having 20 1's randomly distributed in the bitstream (Fig. 2e), corresponding to a bias ratio p = 0.4. Here the bias ratio of the bitstream is defined as the ratio of the number of 1s to the total stream length. The bias ratio in this random bitstream can be controlled by the pulse width and voltage used. For example, the bias ratio changes to p = 0.76 if the pulse width is increased to 1000 ms (Fig. 2f). It is also important to note here we can control only the bias ratio of the bit-stream (i.e. the overall number of 1s), while the locations of individual 1's are random. The randomness is an essential requirement for applications such as bitstream based stochastic computing, where correlation between the locations of 1s can lead to systematic computing errors (24).

Additionally, the probability of switching is related only to the voltage amplitude and the *total time* the voltage is applied, i.e. the switching probability is cumulative. This provides an analog component (e.g. the cumulative time during which the programming voltage is applied) in these binary devices if they are used as synapses in neuromorphic systems (25). Further, the time variable may be discretized such that the probability of switching can in turn be written as a function of the number of (short) pulses of a fixed voltage, regardless of the gap between the pulses. This has the advantage of allowing a look-up table to be used to determine the number of pulses needed to achieve a given probability, rather than more complex timing circuitry. As a proof of concept, we apply a string of pulses, each of which having a 2.5 V amplitude and 100 ms duration, and count the number of pulses it takes for the device to switch. As expected, the distribution for the number of pulses needed to switch the device also follows a Poisson distribution and can be fitted with just one fitting parameter $\tau$. The time constant $\tau$ obtained by counting the number of discrete pulses agrees well with the time constant obtained from applying continuous voltage biases (Fig. 1a). Significantly, almost identical $\tau$ values were obtained when increasing the gap between the applied pulses from 10 ms (Fig 3a) to 500 ms (Fig. 3b) while keeping the pulse width constant, proving the cumulative effect of the programming pulses on switching.

A possible application for the stochastic switching behavior is stochastic computing. Stochastic computing was first proposed in the 1960s (26,27) but like many concepts proposed ahead of their time it never become mainstream due to the exponential growth of digital computers. In stochastic computing, analog values are represented as probabilities in bitstreams. For example, a bitstream B containing 25 percent 1s and 75 percent 0s denotes the number b = 0.25. The length or the structure of B need not be fixed. For example {1,0,0,0}, {0,0,1,0} and {0,0,0,0,1,0,1,0} all are possible representations of b = 0.25. This value depends only on the ratio of the total number of 1s to the total length of the bitstream and not on the value of a particular position in the bitstream. This stochastic representation has an inherent advantage over the binary radix representation when it comes to noise tolerance, since a bit flip only causes an error of $1/n$ where $n$ is the length of the bitstream regardless of where the error occurs, while in the binary radix representation a bit flip can cause an error of $2^{n-1}$ if it occurs at the most significant bit (28). Another advantage of stochastic computing is that relatively simple circuits can be used for tasks which otherwise are computationally more intensive in binary radix representation. For example, the multiplication of two numbers $x$ and $y$ can be achieved by simply performing a bit-wise AND operation on the two bitstreams representing $x$ and $y$ (28). This operation assumes that $x$ and $y$ are independent bitstreams that are completely uncorrelated. Any correlation degrades the accuracy of stochastic computing. For example, if we multiply two identical bitstreams using the AND gate, the product will be $x$, instead of $x^2$. The independence assumption requires the bitstreams to be randomized and in conventional CMOS approaches this is achieved via a stochastic number generator (SNG) (24). SNGs account for the majority of the resource usage in stochastic computing, with a cost as high as 80% of the entire system (24). The ability to generate random bitstreams using memristive devices can represent an intriguing option for efficient computation of data intensive problems such as image processing using stochastic approaches.

In this approach, the bitstreams representing an analog number $x$ is achieved by a memristive device programmed with a predetermined voltage and pulse width to yield random 1s in time domain with a known bias ratio. To modulate this bias ratio or, in other words, to generate another number $y$, all we need to do is to change the programming time or the programming voltage. Bitstreams generated with 2.5 V/300 ms (Fig. 2e) and 2.5 V/1000 ms (Fig. 2f) clearly represent two distinct values, namely, 0.4 and 0.76 respectively. To verify the ability to generate biased bitstreams in multiple devices simultaneously, identical voltage pulses were applied to different devices, shown in Fig. 4a. Again if we digitize the currents by setting 0.1 µA as the threshold, 4 20-bit random bitstreams with similar bias ratios were obtained in the 4 different devices. The spatial variations were minimized here by carefully controlling the fabrication (line edge roughness of the electrodes, thickness and microstructure of the amorphous silicon matrix, etc.), the initial "forming" cycle and the erase process. In particular, in these devices the filament characteristics can be controlled during programming by external circuit parameters such as series-resistance (29) and programming pulse height/width (14,17) so similar switching characteristics can be obtained in different devices.

In Fig. 4 we see that each bit in the bitstream was generated at a different time instance. In other words, the random bits were produced in serial fashion one after the other. This is the commonly termed as "stochastic bitstream in time". For more efficient computing, bitstreams in space domain, i.e. bits generated in parallel or "stochastic bitstreams in space" are also required, which leads to parallel processing of the bitstreams. Here we show that stochastic bitstreams in space can also be obtained by using memristive devices in parallel in an array form.

Bits produced in different physical locations (cells) were generated by programming an array of 4 devices (A, B, C, D) with a single pulse in parallel as shown in Fig. 5a (Inset). Here the state of each device (1 or 0) represents the bits of the bitstream in space and was read out individually after the programming pulse.

As shown in Fig. 5, random combinations can be obtained from the array in different tests, even though an identical programming pulse was applied. In other words, each time the array of 4 devices is programmed, we get a 4-bit wide random bitstream from the array, and exactly which device will switch cannot be predicted at the start of the programming cycle. Besides being used as bitstreams in space in stochastic computing, the 4-device array, when used as a whole as a single synapse, represents a 2-bit weight synapse. The choice of binary synapse or multi-bit synapse in neuromorphic computing has been under long debate. Binary synapses are easier to fabricate, more robust and offer larger dynamic range (4,5). However, multi-bit synapses can lead to more complex system functions and are more desirable. It is possible to achieve multi-bit in binary memristive devices by controlling the programming current (5), however, precise current control in a large network is extremely challenging (30). The concept of using an array of memristive devices to represent a multi-bit synapse was proposed a few years ago (25), and to our knowledge this is the first experimental demonstration of such a concept.

A question can then be raised as to why the device array did not behave as a single, larger device simply having four times the area. In the latter case, since only a dominant filament exists we should only observe one device in ON state in the 4-device array instead. The answer can be obtained after carefully examining the filament formation process. During programming, after the 1$^{st}$ filament bridges the electrodes the voltage drop across the device is subsequently reduced, an effect caused by the voltage divider effect with the external circuit components (i.e. series resistance) (14,17). The reduction of bias voltage in turn slows the growth of additional filaments and results in a single, dominant filament. Indeed, if the 4 devices share the same series resistor, we indeed observed only one device in the ON state after each programming pulse, *i.e.* the device array simply behaves as a single, larger device. However, if each device has its own local, series resistor, as shown in Fig. 5a inset, the completion of the dominant filament in one device does not cause the decrease of voltage seen by other devices thus multiple devices can be switched.

In conclusion, we show binary memristive devices can exhibit native stochastic nature of resistive switching. Even for a fixed voltage on the same device, the wait time is not fixed and is random and broadly distributed. However, the probability of switching can be predicted and controlled by the applied voltage and the pulse width used to program the device. The memristive devices have been used to generate random bitstreams with predicable bias ratios in time and space domains. The array of binary memristive devices can also act as a multi-bit synapse in neuromorphic applications. Rather than trying to eliminate the non-determinism to use these devices in deterministic memory applications, the ability to produce random bitstreams and multi-bit synapses using binary memristive devices based on the native stochastic switching principle may potentially lead to novel non von-Neumann, alternative computing paradigms.

Finally, we would like to point out that this study presents a proof-of-concept demonstration in stochastic and neuromorphic computing approaches by taking advantage of the inherent temporal variations in resistive devices. To implement these concepts in large scale systems, issues like spatial variations, power, speed, endurance and reliability of the devices need to be addressed in subsequent studies. For example, applications in logic require that the devices have very high endurance and the switching probability not degrade with time and be uniform among devices. Recent studies have already shown that similar devices can be very reliable, robust and switch within nanoseconds (31,32,33). Since stochastic operations require lower voltages than conventional binary switching it is expected that even better endurance can be obtained using the proposed approach. Additionally, efficient design of the peripheral circuitry to account for endurance limits under low voltage operations and degradation of switching probability with time need to be carried out in future studies.

METHODS

Experiments were conducted on two-terminal cross-point devices. Device fabrication begins with a blanket deposition of 60nm Boron-doped poly-silicon (with sheet resistance of 1000 ohm / square) on a Si/SiO2 substrate with 100nm $SiO_2$ using a low pressure chemical vapor deposition furnace. Poly-silicon bottom electrodes are patterned by electron beam lithography and reactive ion etching (RIE). Next, a 35nm insulating film of amorphous silicon is deposited in a plasma enhanced chemical vapor deposition system. Silver top electrodes are patterned using electron beam lithography and conventional liftoff method. Subsequently, bottom electrode contacts are opened using photo-lithography and RIE. Finally, large gold contact pads are formed to ensure good electrical connectivity and to provide a suitable surface for wire bonding. Electrical data was collected using a National Instruments data acquisition system (NI USB-6259 BNC) in conjunction with a DL 1211 current preamplifier from DL Industries. Custom code written in MATLAB is used to generate and collect the signals. For automatic testing of multiple cells in parallel, a custom circuit was built using analog switches (ADG1412 from Analog Devices) and a microcontroller (Atmega32u4). The switches were closed during the application of the programming pulse so the 4 devices can be programmed in parallel. During read the switches were sequentially closed so the device states can be read out individually.

ACKNOWLEDGMENT

The authors thank Dr. Zhengya Zhang, Phil Knag and Lin Chen for useful discussions. This work was supported in part by the AFOSR through MURI grant FA9550-12-1-0038 and by the National Science Foundation (NSF) through grant CCF-1217972. This work used the Lurie Nanofabrication Facility at the University of Michigan, a member of the National Nanotechnology Infrastructure Network (NNIN) funded by NSF.

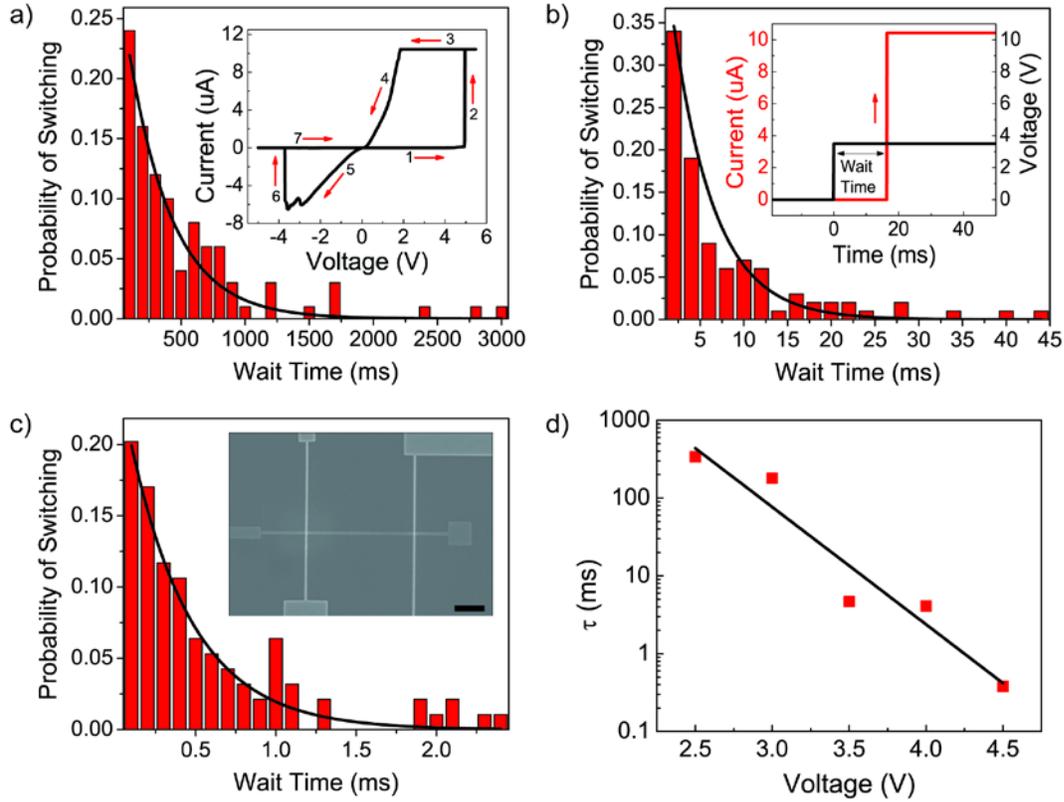

Fig. 1. Random wait time distribution. (a-c) Distributions of wait times for applied voltages of 2.5 V (a), 3.5 V (b) and 4.5 V (c). Solid lines: Fitting to the Poisson distribution of Eq. 1 using t as the only fitting parameter. τ = 340 ms, 4.7 ms and 0.38 ms for a-c, respectively. Insets: a: DC switching curves, b: Example of a wait time measurement and c: Scanning electron micrograph of a typical device (scale bar: 2.5 μm). (d) Dependence of t on the programming voltage. Solid squares were obtained from fitting of the wait time distributions while the solid line is an exponential fit.

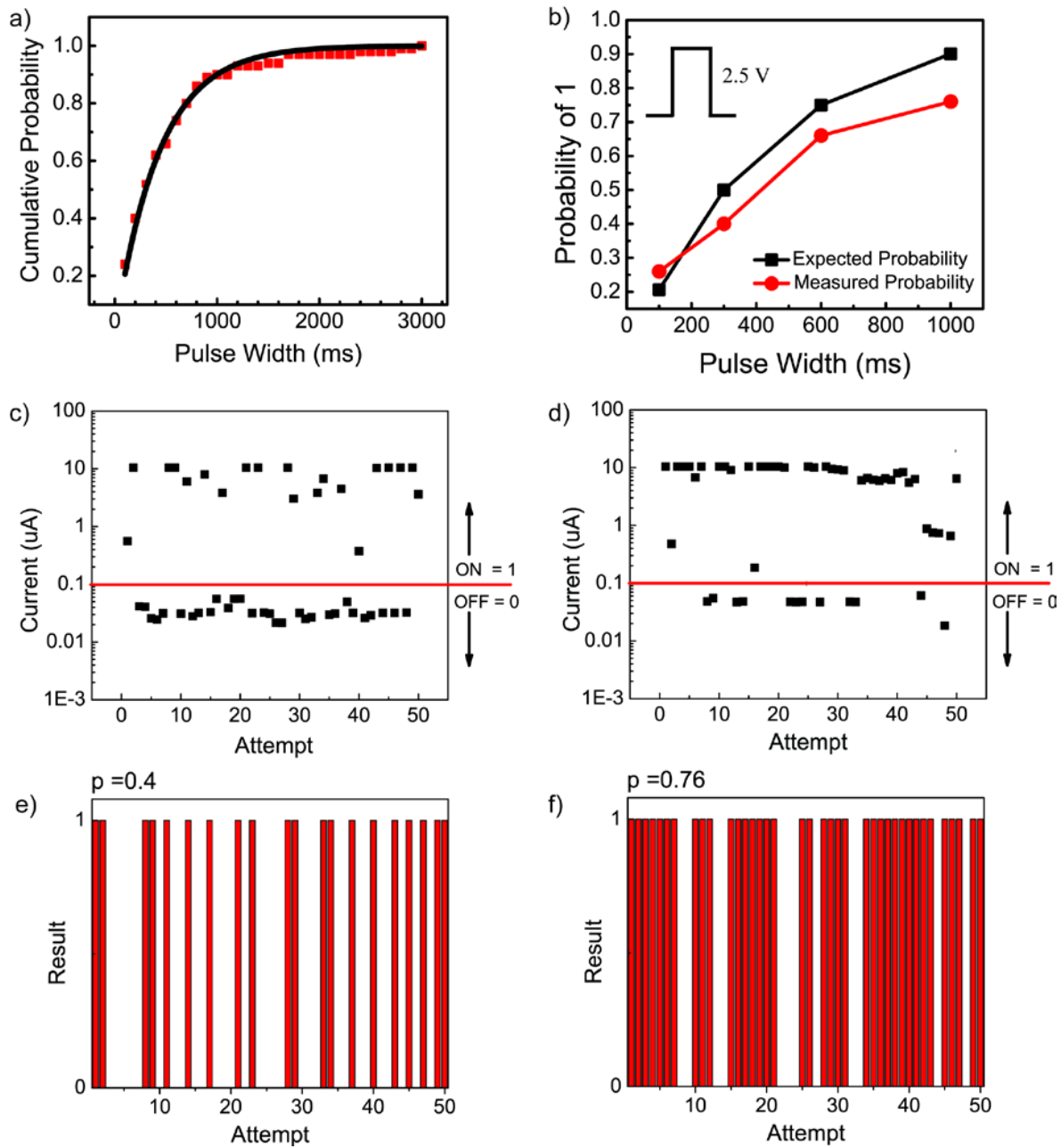

Fig. 2. Probability of switching within a single pulse. (a) Pulse width dependence. Solid line shows values predicted from Eq. (2), squares were obtained from measuring the cumulative probability obtained from Fig. 1a. (b) Expected and measured probability using a single 2.5 V pulse. (c-d) Device current measured after repeated application of a single 2.5 V, 300 ms (c) and 1000 ms (d) pulse. The device was reset after each measurement. (e-f) Corresponding bitstreams of (e) p = 0.4 and (f) p = 0.76 corresponding to (c) and (d).

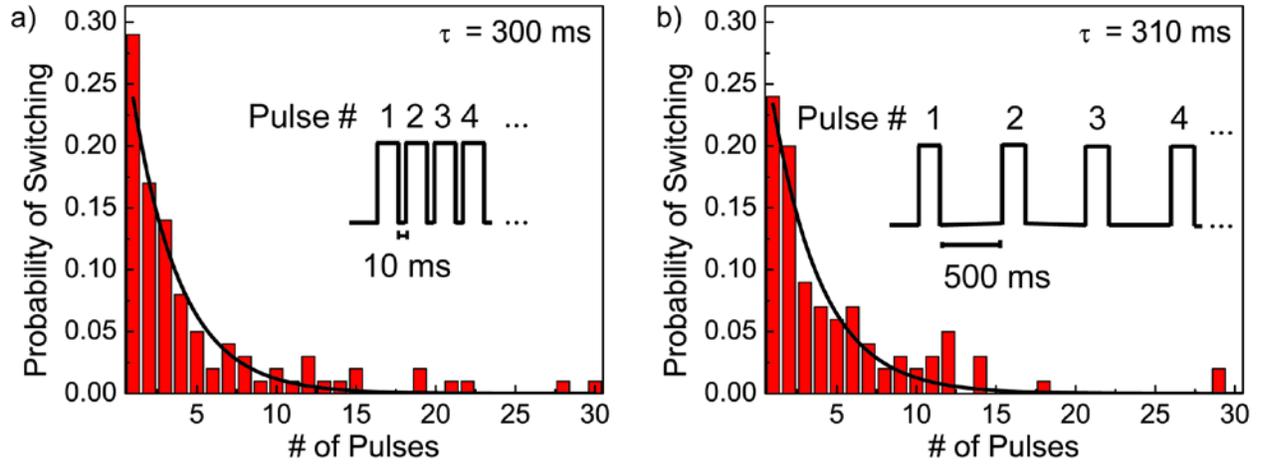

Fig. 3. Switching probability when subjected to a series of short pulses. The average switching time can be calculated by measuring the number of pulses, regardless of the gap between the pulses. The pulse width was kept constant at 100 ms while the gap was changed from 10 ms (a) to 500 ms (b). The voltage amplitude was fixed at 2.5V for all pulses.

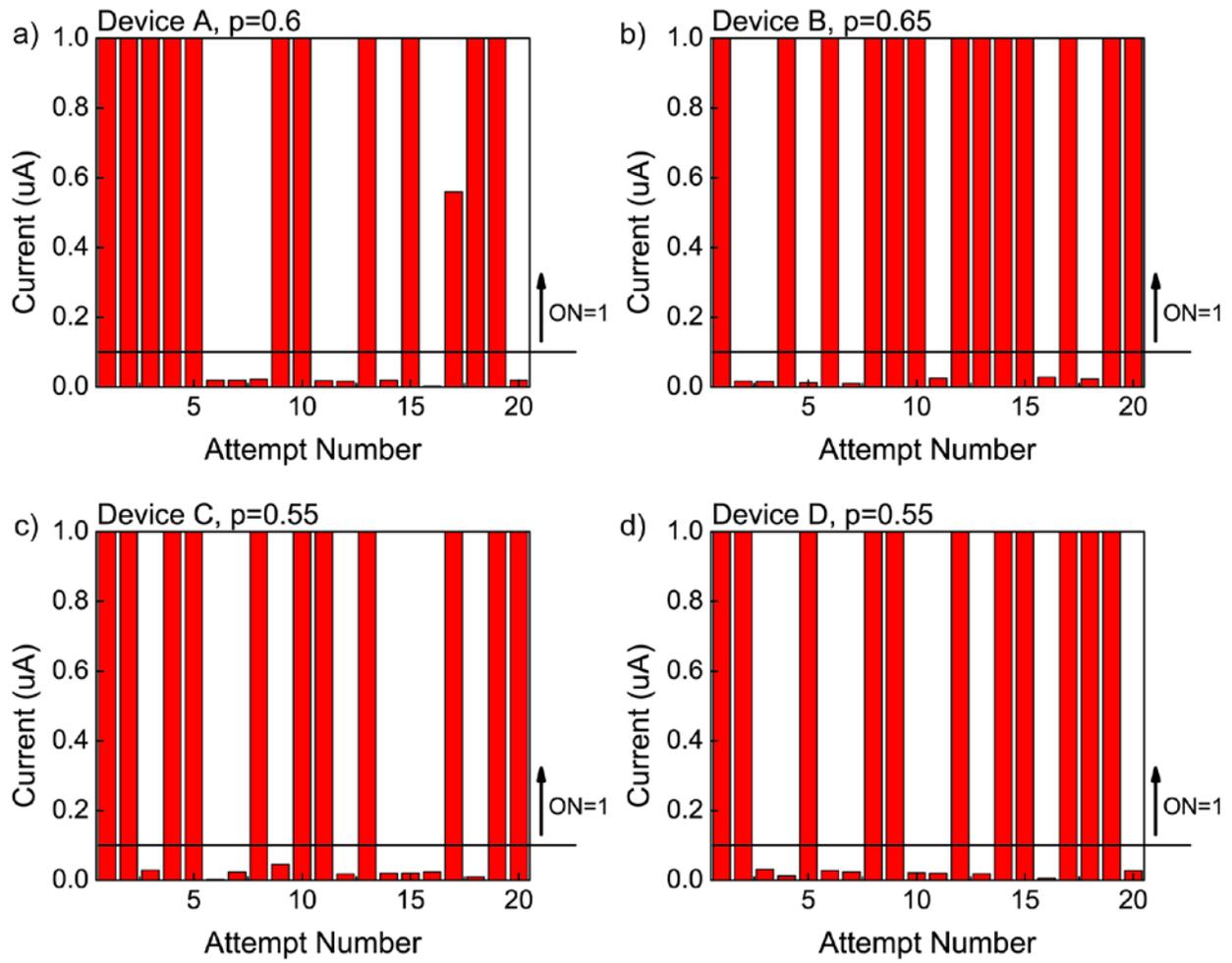

Fig. 4. Stochastic switching in different devices (a-d) obtained from a second fabrication run. The devices were measured with twenty 2.75 V, 1000 ms pulses. The devices were erased after each measurement.

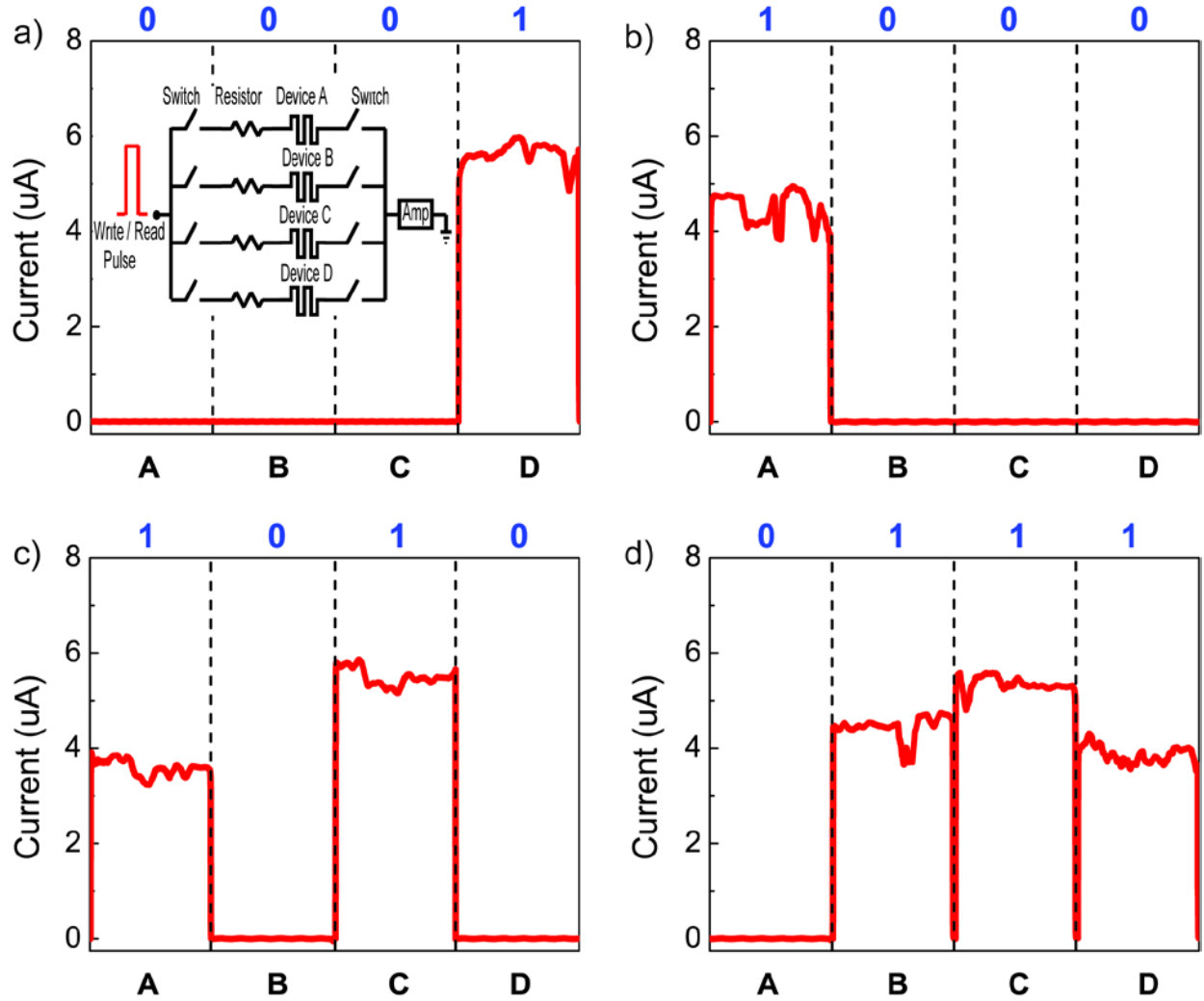

Fig. 5. Stochastic programming of a device array. 4 representative combinations {0001} (a) {1000} (b) {1010} (c) and {0111} (d) are shown here out of a total of $2^4=16$ combinations. The different combinations were obtained in the same array under identical pulses. The devices were programmed in parallel using a single 2.75 V, 1000 ms pulse and their states were measured using 2 V, 100 ms pulses individually after the programming pulse. The array was reset after each measurement. Inset to (a) shows the circuit used for this study.